\documentclass[magazine]{IEEEtran}

\usepackage[T1]{fontenc}
\usepackage[utf8]{inputenc}
\usepackage[super,comma,sort&compress]{natbib}
\usepackage{amsmath, mathtools}
\usepackage{amsfonts,amsthm,bm}
\usepackage{amssymb}
\usepackage{comment}
\usepackage{graphicx}
\usepackage{standalone}
\usepackage{dsfont}
\usepackage{mathtools}
\usepackage{color, colortbl}
\usepackage{soul}
\usepackage[normalem]{ulem}
\usepackage[acronym,shortcuts]{glossaries}
\usepackage{arydshln}
\usepackage{bbm}
\usepackage{svg} 
\usepackage{algorithm,algorithmic}
\usepackage{adjustbox}
\usepackage[inline]{enumitem}
\usepackage{multirow}
\usepackage{tabularx}
\usepackage{booktabs}

\usepackage{comment}
\usepackage{tikz}
\usepackage{pgfplots}
\pgfplotsset{compat=newest}
\pgfplotsset{plot coordinates/math parser=false}
\newlength\fheight
\newlength\fwidth
\usetikzlibrary{plotmarks,patterns,decorations.pathreplacing,backgrounds,calc,arrows,arrows.meta,spy,matrix}
\usepgfplotslibrary{patchplots,groupplots}
\usepackage{tikzscale}
\usepackage{balance}
\usepackage{stfloats} 

\usepackage{hyperref}
\usepackage[capitalize]{cleveref}
\crefname{section}{Sec.}{Secs.}

\usepackage{subcaption}
\usepackage{graphicx}
\usepackage{caption}
\usepackage{xcolor}
\definecolor{ieeeblue}{HTML}{00629B}
\definecolor{ieeeorange}{HTML}{FFA300}
\definecolor{ieeegreen}{HTML}{78BE20}
\definecolor{ieeered}{HTML}{BA0C2F}

\usetikzlibrary {patterns.meta}

\usetikzlibrary{patterns}

\usepackage{ulem}

\providecolor{added}{rgb}{0,0,1}
\providecolor{deleted}{rgb}{1,0,0}

\makeatletter
\newcommand{\new}[1]{{\textcolor{blue}{#1}}}

\makeatletter
\newcommand\remembertext[2]{
  \immediate\write\@auxout{\unexpanded{\global\long\@namedef{mytext@#1}{#2}}}%
  {\color{blue} #2}%
}
\newcommand\recalltext[1]{%
  \new{\ifcsname mytext@#1\endcsname
    \fontsize{10.5}{12.5}\selectfont\@nameuse{mytext@#1}%
  \else
    ``??''
  \fi
}}
\makeatother

\makeatletter
\renewcommand{\IEEEauthorrefmark}[1]{\textsuperscript{#1}}
\makeatother

\newacronym{3gpp}{3GPP}{3rd Generation Partnership Project}
\newacronym{adc}{ADC}{Analog to Digital Converter}
\newacronym{pdr}{PDR}{Packet Delivery Ratio}
\newacronym{5g}{5G}{5th generation}
\newacronym{6g}{6G}{6th generation}
\newacronym{ai}{AI}{Artificial Intelligence}
\newacronym{aimd}{AIMD}{Additive Increase Multiplicative Decrease}
\newacronym{am}{AM}{Acknowledged Mode}
\newacronym{tn}{TN}{Terrestrial Network}
\newacronym{amc}{AMC}{Adaptive Modulation and Coding}
\newacronym{aqm}{AQM}{Active Queue Management}
\newacronym{awgn}{AGWN}{Additive White Gaussian Noise}
\newacronym{balia}{BALIA}{Balanced Link Adaptation}
\newacronym{bdp}{BDP}{Bandwidth-Delay Product}
\newacronym{bf}{BF}{beamforming}
\newacronym{uu}{Uu}{Universal um}
\newacronym{refp}{RP}{Reference Point}
\newacronym{cc}{CC}{Congestion Control}
\newacronym{cdf}{CDF}{Cumulative Distribution Function}
\newacronym{cn}{CN}{Core Network}
\newacronym{cqi}{CQI}{Channel Quality Information}
\newacronym{cp}{CP}{Control Plane}
\newacronym{up}{UP}{User Plane}
\newacronym{upf}{UPF}{User Plane Function}
\newacronym{csirs}{CSI-RS}{Channel State Information - Reference Signal}
\newacronym{sib}{SIB}{System Information Block}
\newacronym{dc}{DC}{Dual Connectivity}
\newacronym{rb}{RB}{Resource Block}
\newacronym{dce}{DCE}{Direct Code Execution}
\newacronym{dci}{DCI}{downlink control onformation}
\newacronym{udp}{UDP}{User Datagram Protocol}
\newacronym{dl}{DL}{downlink}
\newacronym{drl}{DRL}{deep reinforcement learning}
\newacronym{fcfs}{FCFS}{first-come-first-served}
\newacronym{dmr}{DMR}{Deadline Miss Ratio}
\newacronym{fspl}{FSPL}{free-space path loss}
\newacronym{eirp}{EIRP}{Effective Isotropic Radiated Power}
\newacronym{dmrs}{DMRS}{DeModulation Reference Signal}
\newacronym{e2e}{E2E}{End-to-End}
\newacronym{ppp}{PPP}{Poission Point Process}
\newacronym{aoi}{AoI}{Area of Interest}
\newacronym{cpu}{CPU}{Central Processing Unit}
\newacronym{gpu}{GPU}{Graphics Processing Unit}
\newacronym{tpu}{TPU}{Tensor Processing Unit}
\newacronym{ta}{TA}{Timing Advance}
\newacronym{si}{SI}{Study Item}
\newacronym{ecn}{ECN}{Explicit Congestion Notification}
\newacronym{edf}{EDF}{Earliest Deadline First}
\newacronym{enb}{eNB}{eNodeB}
\newacronym{epc}{EPC}{Evolved Packet Core}
\newacronym{es}{ES}{Edge Server}
\newacronym{cav}{CAV}{Connected and Autonomous Vehicle}
\newacronym{fdma}{FDMA}{Frequency Division Multiple Access}
\newacronym{fdd}{FDD}{Frequency Division Duplexing}
\newacronym{tdm}{TDM}{Time Division Multiplexing}
\newacronym{upa}{UPA}{Uniform Planar Array}
\newacronym{car}{CAR}{Circular Aperture Reflector }
\newacronym[firstplural=Radio Access Technologies (RATs)]{rat}{RAT}{Radio Access Technology}
\newacronym[firstplural=Radio Access Technology (RTs)]{rt}{RT}{Radio Technology}
\newacronym{fs}{FS}{Fast Switching}
\newacronym{isd}{ISD}{inter-site distance}
\newacronym{ftp}{FTP}{File Transfer Protocol}
\newacronym{gnb}{gNB}{Next Generation NodeB}
\newacronym{harq}{HARQ}{Hybrid Automatic Repeat reQuest}
\newacronym{hetnet}{HetNet}{Heterogeneous Network}
\newacronym{hh}{HH}{Hard Handover}
\newacronym{hol}{HOL}{Head-of-Line}
\newacronym{ia}{IA}{Initial Access}
\newacronym{imt}{IMT}{International Mobile Telecommunication}
\newacronym{embb}{eMBB}{enhanced Mobile BroadBand}
\newacronym{emtc}{eMTC}{enhanced Machine-Type Communication}
\newacronym{iot}{IoT}{Internet of Things}
\newacronym{los}{LOS}{Line of Sight}
\newacronym{lte}{LTE}{Long Term Evolution}
\newacronym{m2m}{M2M}{Machine to Machine}
\newacronym{mac}{MAC}{Medium Access Control}
\newacronym{mc}{MC}{Multi-Connectivity}
\newacronym{mcs}{MCS}{Modulation and Coding Scheme}
\newacronym{mec}{MEC}{Mobile Edge Cloud}
\newacronym{mi}{MI}{Mutual Information}
\newacronym{mimo}{MIMO}{Multiple Input Multiple Output}
\newacronym{mmwave}{mmWave}{millimeter wave}
\newacronym{mptcp}{MP-TCP}{Multipath TCP}
\newacronym{mr}{MR}{Maximum Rate}
\newacronym{mss}{MSS}{Maximum Segment Size}
\newacronym{mtd}{MTD}{Machine-Type Device}
\newacronym{mtu}{MTU}{Maximum Transmission Unit}
\newacronym{nfv}{NFV}{Network Function Virtualization}
\newacronym{vnf}{VNF}{Virtualization Network Function}
\newacronym{gv}{GV}{ground vehicle}
\newacronym{vec}{VEC}{Vehicular Edge Computing}
\newacronym{dn}{DN}{Data Network}
\newacronym{sdn}{SDN}{Software Defined Networking}
\newacronym{nlos}{NLOS}{Non Line of Sight}
\newacronym{nlosb}{NLOSb}{Building Non Line of Sight}
\newacronym{nlosv}{NLOSv}{Vehicle Non Line of Sight}
\newacronym{nr}{NR}{New Radio}
\newacronym{ofdm}{OFDM}{Orthogonal Frequency Division Multiplexing}
\newacronym{pdcch}{PDCCH}{Physical Downlonk Control Channel}
\newacronym{sctp}{SCTP}{Stream Control Transport Protocol}
\newacronym{sdap}{SDAP}{Service Data Adaptation Protocol}
\newacronym{pdcp}{PDCP}{Packet Data Convergence Protocol}
\newacronym{pdsch}{PDSCH}{Physical Downlink Shared Channel}
\newacronym{pdu}{PDU}{Packet Data Unit}
\newacronym{pf}{PF}{Proportional Fair}
\newacronym{pgw}{PGW}{Packet Gateway}
\newacronym{sgw}{SGW}{Serving Gateway}
\newacronym{phy}{PHY}{Physical}
\newacronym{pbch}{PBCH}{Physical Broadcast Channel}
\newacronym[plural=\gls{mme}s,firstplural=Mobility Management Entities (MMEs)]{mme}{MME}{Mobility Management Entity}
\newacronym{prb}{PRB}{Physical Resource Block}
\newacronym{pss}{PSS}{Primary Synchronization Signal}
\newacronym{pucch}{PUCCH}{Physical Uplink Control Channel}
\newacronym{pusch}{PUSCH}{Physical Uplink Shared Channel}
\newacronym{rach}{RACH}{Random Access Channel}
\newacronym{ran}{RAN}{Radio Access Network}
\newacronym{red}{RED}{Random Early Detection}
\newacronym{rf}{RF}{Radio Frequency}
\newacronym{rlc}{RLC}{Radio Link Control}
\newacronym{rlf}{RLF}{Radio Link Failure}
\newacronym{rrc}{RRC}{Radio Resource Control}
\newacronym{rrm}{RRM}{Radio Resource Management}
\newacronym{rr}{RR}{Round Robin}
\newacronym{rs}{RS}{Remote Server}
\newacronym{rsrp}{RSRP}{Reference Signal Received Power}
\newacronym{rss}{RSS}{Received Signal Strength}
\newacronym{rtt}{RTT}{Round Trip Time}
\newacronym{rw}{RW}{Receive Window}
\newacronym{rx}{RX}{Receiver}
\newacronym{sa}{SA}{standalone}
\newacronym{sack}{SACK}{Selective Acknowledgment}
\newacronym{sap}{SAP}{Service Access Point}
\newacronym{sch}{SCH}{Secondary Cell Handover}
\newacronym{scoot}{SCOOT}{Split Cycle Offset Optimization Technique}
\newacronym{sdma}{SDMA}{Spatial Division Multiple Access}
\newacronym{sinr}{SINR}{Signal to Interference plus Noise Ratio}
\newacronym{sm}{SM}{Saturation Mode}
\newacronym{snr}{SNR}{Signal-to-Noise Ratio}
\newacronym{son}{SON}{Self-Organizing Network}
\newacronym{ss}{SS}{Synchronization Signal}
\newacronym{srs}{SRS}{Sounding Reference Signal}
\newacronym{sss}{SSS}{Secondary Synchronization Signal}
\newacronym{tb}{TB}{Transport Block}
\newacronym{tcp}{TCP}{Transmission Control Protocol}
\newacronym{tdd}{TDD}{Time Division Duplexing}
\newacronym{tdma}{TDMA}{Time Division Multiple Access}
\newacronym{tfl}{TfL}{Transport for London}
\newacronym{tm}{TM}{Transparent Mode}
\newacronym{prr}{PRR}{Packet Reception Ratio}
\newacronym{trp}{TRP}{Transmitter Receiver Pair}
\newacronym{tti}{TTI}{Transmission Time Interval}
\newacronym{ttt}{TTT}{Time-to-Trigger}
\newacronym{tx}{TX}{Transmitter}
\newacronym{ue}{UE}{User Equipment}
\newacronym{ul}{UL}{uplink}
\newacronym{uml}{UML}{Unified Modeling Language}
\newacronym{um}{UM}{Unacknowledged Mode}
\newacronym{utc}{UTC}{Urban Traffic Control}
\newacronym{vm}{VM}{Virtual Machine}
\newacronym{rsrq}{RSRQ}{Reference Signal Received Quality}
\newacronym{rssi}{RSSI}{Received Signal Strength Indicator}
\newacronym{crs}{CRS}{Cell Reference Signal}
\newacronym{v2v}{V2V}{Vehicle-to-Vehicle}
\newacronym{v2i}{V2I}{Vehicle-to-Infrastructure}
\newacronym{v2n}{V2N}{Vehicle-to-Network}
\newacronym{v2x}{V2X}{Vehicle-to-Everything}
\newacronym{vn}{VN}{Vehicular Node}
\newacronym{dsrc}{DSRC}{Dedicated Short Range Communication}
\newacronym{ci}{CI}{context information}
\newacronym{voi}{VoI}{value of information}
\newacronym{gps}{GPS}{Global Positioning System}
\newacronym{qos}{QoS}{Quality of Service}
\newacronym{qoe}{QoE}{Quality of Experience}
\newacronym{ml}{ML}{Machine Learning}
\newacronym{ahp}{AHP}{Analytic Hierarchy Process}
\newacronym{lidar}{LIDAR}{Light Detection and Ranging}
\newacronym{sumo}{SUMO}{Simulation of Urban MObility}
\newacronym{wave}{WAVE}{Wireless Access in Vehicular Environment}
\newacronym{c-its}{C-ITS}{Connected Intelligent Transportation System}
\newacronym{dash}{DASH}{Dynamic Adaptive Streaming over HTTP}
\newacronym{http}{HTTP}{HyperText Transfer Protocol}
\newacronym{nt}{NT}{Non-Terrestrial}
\newacronym{ntc}{NTC}{non-terrestrial communication}
\newacronym{ntn}{NTN}{Non-Terrestrial Network}
\newacronym{haps}{HAPS}{High Altitude Platform Station}
\newacronym{hap}{HAP}{High Altitude Platform}
\newacronym{leo}{LEO}{Low Earth Orbit}
\newacronym{meo}{MEO}{Medium Earth Orbit}
\newacronym{geo}{GEO}{Geostationary Earth Orbit}
\newacronym{uav}{UAV}{Unmanned Aerial Vehicle}
\newacronym{nsat}{nSAT}{Nanosatellite}
\newacronym{ehf}{EHF}{extremely high-frequency}
\newacronym{ioe}{IoE}{Internet of Everyone}
\newacronym{gan}{GaN}{Gallium Nitride}
\newacronym{af}{AF}{amplify-and-forward}
\newacronym{csi}{CSI}{channel state information}
\newacronym{ecdf}{ECDF}{empirical cumulative distribution function}
\newacronym{f}{F}{flexible}
\newacronym{fpga}{FPGA}{field programmable gate array}
\newacronym{fov}{FoV}{field-of-view}
\newacronym{km}{KM}{K-means}
\newacronym{kmed}{KMed}{K-medoids}
\newacronym{iab}{IAB}{Integrated Access and Backhaul}
\newacronym{bap}{BAP}{backhaul adaptation protocol}
\newacronym{irs}{IRS}{intelligent reflecting surface}
\newacronym{lsfc}{LSFC}{large-scale fading coefficient}
\newacronym{noma}{NOMA}{non-orthogonal multiple access}
\newacronym{fdm}{FDM}{frequency-division multiplexing}
\newacronym{sdm}{SDM}{space-division multiplexing}
\newacronym{ofdma}{OFDMA}{orthogonal frequency-division multiple access}
\newacronym{oma}{OMA}{orthogonal multiple access}
\newacronym{isl}{ISL}{Inter-Satellite Links}
\newacronym{rsma}{RSMA}{rate-splitting multiple access}
\newacronym{scm}{SCM}{spatial channel model}
\newacronym{siso}{SISO}{single input single output}
\newacronym{svd}{SVD}{singular value decomposition}
\newacronym{5gc}{5GC}{5G Core}
\newacronym{thz}{THz}{Terahertz}
\newacronym{ula}{ULA}{uniform linear array}
\newacronym{uma}{UMa}{urban macro-cell}
\newacronym{umi}{UMi}{urban micro-cell}
\newacronym{mt}{MT}{mobile terminal}
\newacronym{cu}{CU}{centralized unit}
\newacronym{du}{DU}{distributed unit}
\newacronym{dag}{DAG}{directed acyclic graph}
\newacronym{st}{ST}{spanning tree}
\newacronym{rma}{RMa}{rural macrocell}
\newacronym{inf}{InF}{indoor factory}
\newacronym{ngc}{NGC}{next generation core}
\newacronym{gtp}{GTP}{GPRS Tunnelling Protocol}
\newacronym{tft}{TFT}{Traffic Flow Template}
\newacronym{teid}{TEID}{Tunnel Endpoint Identifier}
\newacronym{tnl}{TNL}{Transport Network Layer}
\newacronym{amf}{AMF}{Access and Mobility Management Function}
\newacronym{ngso}{NGSO}{Non-Geostationary Orbit}
\newacronym{redcap}{RedCap}{Reduced Capability}
\newacronym{ng}{NG}{Next Generation}
\newacronym{fr1}{FR1}{Frequency Range 1}
\newacronym{fr2}{FR2}{Frequency Range 2}
\newacronym{prach}{PRACH}{Physical Random Access Channel}
\newacronym{ro}{RO}{RACH Occasion}
\newacronym{app}{APP}{Application}
\newacronym{vsat}{VSAT}{Very Small Aperture Terminal}
\newacronym{ack}{ACK}{Acknowledgment}
\newacronym{cwnd}{CWND}{Congestion Window}
\newacronym{rar}{RAR}{Random Access Response}
\newacronym{n6}{N6}{Network 6}
\newacronym{rto}{RTO}{Retransmission Timeout}
\newacronym{ims}{IMS}{IP Multimedia Subsystem}
\newacronym{6gr}{6GR}{6G Radio}
\newacronym{gp}{GP}{Guard Period}

\title{5G NR Non-Terrestrial Networks: \\From Early Results to the Road Ahead}

\author{Mattia Figaro\IEEEauthorrefmark{1}, Francesco Rossato\IEEEauthorrefmark{1},~\IEEEmembership{Student Members, IEEE}, Marco~Giordani\IEEEauthorrefmark{1},~\IEEEmembership{Senior Member, IEEE}, \\ Alessandro Traspadini\IEEEauthorrefmark{1},~\IEEEmembership{Student Member,~IEEE}, Takayuki Shimizu\IEEEauthorrefmark{2}, Chinmay Mahabal\IEEEauthorrefmark{2}, Sanjeewa Herath\IEEEauthorrefmark{2}, Chunghan Lee\IEEEauthorrefmark{2}, Dogan Kutay Pekcan\IEEEauthorrefmark{2}, Michele Zorzi\IEEEauthorrefmark{1},~\IEEEmembership{Fellow, IEEE}
\thanks{\IEEEauthorrefmark{1} Department of Information Engineering, University of Padova, Padova, Italy. 
Emails: \{mattia.figaro, francesco.rossato, marco.giordani, alessandro.traspadini, michele.zorzi\}@dei.unipd.it.}
\thanks{\IEEEauthorrefmark{2} R\&D InfoTech Labs, Toyota Motor North America Inc., USA. Email: \{takayuki.shimizu, chinmay.mahabal, sanjeewa.herath, chunghan.lee1, dogan.pekcan\}@toyota.com}
\thanks{The corresponding author is Francesco Rossato (email francesco.rossato.6@phd.unipd.it).}
}
\begin{document}
\maketitle

\begin{abstract}


This paper overviews the 3GPP 5G NR-NTN standard, detailing the evolution from Rel. 18 to 19 and innovations for Rel. 20. Using realistic ns-3 simulations validated against 3GPP calibration data, we evaluate various satellite network configurations. The results highlight the potential of NTNs to extend wireless connectivity to remote areas, serve requests during emergency, and alleviate terrestrial network congestion.
\end{abstract}

\glsresetall

\begin{tikzpicture}[remember picture,overlay]
\node[anchor=north,yshift=-10pt] at (current page.north) {\parbox{\dimexpr\textwidth-\fboxsep-\fboxrule\relax}{
\centering\footnotesize This paper has been accepted for publication in npj Wireless Technology.\\
Please cite it as: M. Figaro, F. Rossato, M. Giordani, A. Traspadini, T. Shimizu, C. Mahabal, S. Herath, C. Lee, D. Pekcan, M. Zorzi, `5G NR Non-Terrestrial Networks: From Early Results to the Road Ahead`, npj Wireless Communications, 2026.\\ 
}};
\end{tikzpicture}

\section*{Introduction}
\label{sec:introduction}
In recent years, the research community has been exploring the potential of \glspl{ntn}, where satellites, \glspl{uav}, and \glspl{hap} act as aerial/space base stations to extend connectivity on Earth beyond the boundaries of \glspl{tn}~\cite{giordani2020non}.
Specifically, \glspl{uav} operate at low altitudes (below 500 m) and offer flexible, on-demand, though short-term, coverage. \glspl{hap} are in the stratosphere (around 20 km) and provide broad connectivity, but have stabilization and energy constraints.
Satellites, especially in the \glspl{leo}, offer wide-area coverage with lower latency with respect to those at higher altitudes, making them especially appealing for \gls{ntn}, as demonstrated by the many commercial Internet access deployments based on \gls{leo} constellations. Other solutions involve the use of \gls{meo} and \gls{geo} satellites, as a compromise between coverage and latency.

The development of \glspl{ntn} is a foundational pillar in the evolution towards future cellular systems such as 6G~\cite{giordani2020toward, rp251881}, to support ubiquitous, intelligent, and resilient global connectivity.
Notably, \glspl{ntn} enable a broad range of applications across diverse
sectors. They provide standalone broadband connectivity in areas with no or limited terrestrial infrastructure, such as remote regions, deserts, forests, national parks, airspace, and oceans, helping bridge the digital divide~\cite{Chaoub20216g}. 
Furthermore, they can potentially support the communication needs of numerous vertical industries, including: connected vehicles and cooperative traffic systems in the transportation sector; real-time digital twins, automation, and robotics in
manufacturing; massive \gls{iot}, sensors, scalable connectivity, and edge computing in
smart cities and infrastructure; global, seamless connectivity in aerospace;
immersive applications such as holographic learning, AR/VR classrooms, and educational digital twins for both urban and rural areas; real-time trading, secure mobile transactions, and decentralized finance in the financial sector; precision agriculture, \gls{uav}-based field monitoring, and climate data analysis in environmental monitoring; autonomous
delivery and inventory management in retail and logistics; and smart grids and
remote asset management in the energy and utility sectors. 
\glspl{ntn} also enable mission-critical communication and disaster response in public safety and emergency services.

In this context, the \gls{3gpp} has formalized the use of NTN, especially satellites, for communication since Rel. 17, with new Work Items on 5G NTN (or 3GPP 5G \gls{nr}-NTN)~\cite{rp221169}.
Since then, the standard has continued to evolve, with new specifications and innovations toward Rel. 19 and 20, marking an important step toward 6G.
Along these lines, Hosseinian \emph{et al.}~\cite{hosseinian2021review}  provided an early overview of the 3GPP 5G NR-NTN standardization landscape; however, the paper, published in 2021, does not capture the most recent developments leading up to Rel. 20.
More recently, Lin~\cite{lin20253gpp} presented a broad roadmap of the 3GPP standardization timeline from Rel.~15 to Rel.~20. Even though the analysis covers several key technologies such as {MIMO}, AI/ML, Industrial IoT, V2X, ISAC and XR, the discussion on NTN remains limited to just a few paragraphs.
Conversely, Tong \emph{et al.} \cite{tong2025review} focused specifically on the {3GPP} 5G {NR}-{NTN} starting from Rel. 17, although the study remains high-level, and does not provide any quantitative performance evaluation.

In fact, rigorous performance evaluation of 3GPP 5G NR-NTN scenarios remains in the early stages.
For example, Sedin \emph{et al.} in \cite{sedin2020throughput} examined the throughput and capacity of LEO NTN in terms of spectral efficiency, considering different frequency bands and antenna configurations. 
Similarly, Roshdi \emph{et al.} \cite{roshdi2024performance} analyzed the performance of an integrated system combining a LEO satellite with a conventional terrestrial base station in terms of packet loss and data rate.
In our previous work \cite{giordani2020satellite}, we also assessed the feasibility of establishing high-capacity satellite links in the Ka band.
However, most of these studies are primary conceptual or mainly focus on link-layer metrics, impose simplifying assumptions on the protocol stack, or are not fully aligned with the 3GPP 5G NR-NTN specifications.
A notable exception is the work in~\cite{badini2025network}, where the authors presented some full-stack end-to-end simulation results for NTN using ns-3. However, they specifically focused on satellite handover, exploring metrics such as the average number of handovers and the handover rate, but the analysis did not extend to end-to-end performance metrics such as throughput and latency, which is instead the primary objective of our work.


To address these gaps, in Sec. 3GPP 5G NR-NTN Overview we review the most recent \gls{3gpp} standardization activities related to 5G \gls{nr}-\gls{ntn} at the time of this publication. Then, in Sec. End-to-End Evaluation of NTN Scenarios we evaluate the performance of a satellite network in terms of throughput, \gls{pdr}, and latency, through end-to-end system-level simulations in ns-3, thereby considering the effect of the full \gls{3gpp} 5G \gls{nr}-\gls{ntn} stack. Simulations are based on the \gls{3gpp} calibration results and specifications reported in~\cite{38821}, as a function of different configuration parameters.


\begin{table*}[t!]
    \caption{Relevant 3GPP technical reports and specifications on 5G NR-NTN for Rel. 17, 18, 19, 20, developed within the Radio Access Network (RAN) Technical Specification Groups/Working Groups (TSGs/WGs).} 
    \label{tab:3gpp}
\centering
\footnotesize
\renewcommand{\arraystretch}{0.9} 
\begin{tabular}{|p{1.2cm}|p{1cm}|p{12cm}|p{2cm}|}
\hline
{\color[HTML]{252525} \textbf{Rel.}} & {\color[HTML]{252525} \textbf{TSG/WG}} & {\color[HTML]{252525} \textbf{Title}}                                                                                            & {\color[HTML]{252525} \textbf{Documents}}                                                                                                                    \\ \hline
{\color[HTML]{252525} 15}               & {\color[HTML]{252525} RAN}                    & {\color[HTML]{252525} Study on NR to support NTN}                                                           & {\color[HTML]{252525} { TR 38.811}}                                                                                                                       \\ \hline
{\color[HTML]{252525} 16}               & {\color[HTML]{252525} RAN3}                   & {\color[HTML]{252525} Study on solutions for NR to support NTN}                                                                  & {\color[HTML]{252525} { TR 38.821}}                                                                                                                       \\ \hline
{\color[HTML]{252525} 17}               & {\color[HTML]{252525} RAN4}                   & {\color[HTML]{252525} Solutions for NR to support NTN}                                                                           & {\color[HTML]{252525} \begin{tabular}[c]{@{}l@{}}RP-221946\\ TR 38.863\\ TS 38.108\\ TS 38.101-5\\ TS 38.181\end{tabular}}                                   \\ \hline
{\color[HTML]{252525} 17}               & {\color[HTML]{252525} RAN5}                   & {\color[HTML]{252525} UE Conformance - Solutions for NR to support NTN plus CT aspects}                                          & {\color[HTML]{252525} { TS 38.521-5}}                                                                                                                     \\ \hline
{\color[HTML]{252525} 18}               & {\color[HTML]{252525} RAN}                    & {\color[HTML]{252525} Study on self-evaluation towards the IMT-2020 submission of the 3GPP Satellite Radio} & {\color[HTML]{252525} { TR 37.911}}                                                                                                                       \\ \hline
{\color[HTML]{252525} 18}               & {\color[HTML]{252525} RAN}                    & {\color[HTML]{252525} Study on requirements and use cases for network verified UE location for NTN in NR}                         & {\color[HTML]{252525} { TR 38.882}}                                                                                                                       \\ \hline
{\color[HTML]{252525} 18}               & {\color[HTML]{252525} RAN4}                   & {\color[HTML]{252525} Introduction of the satellite L-/S-band for NR}                                                            & {\color[HTML]{252525} { TR 38.741}}                                                                                                                     \\ \hline
{\color[HTML]{252525} 17, 18, 19}               & {\color[HTML]{252525} RAN1}                   & {\color[HTML]{252525} Physical Layer Design for 5G NR}                                                                           & {\color[HTML]{252525} \begin{tabular}[c]{@{}l@{}}TS 38.211\\ TS 38.212\\ TS 38.213\\ TS 38.214\end{tabular}}                                   \\ \hline

{\color[HTML]{252525} 17, 18, 19}               & {\color[HTML]{252525} RAN2}                   & {\color[HTML]{252525} 5G NR MAC and RRC Layer Enhancements and Standardization}                                                                           & {\color[HTML]{252525} \begin{tabular}[c]{@{}l@{}}RP-251954\\ RP-252886\\ TS 38.321\\ TS 38.331\end{tabular}}                                   \\ \hline

{\color[HTML]{252525} 17, 18, 19}               & {\color[HTML]{252525} RAN2}                   & {\color[HTML]{252525} 5G NR Architecture and Mobility Procedures}                                                                           & {\color[HTML]{252525} \begin{tabular}[c]{@{}l@{}}TS 38.300\\ TS 38.304 
\end{tabular}}                                   \\ \hline
{\color[HTML]{252525} 20}               & {\color[HTML]{252525} RAN1}                   & {\color[HTML]{252525} Study on GNSS resilient NR-NTN operation}                                          & {\color[HTML]{252525} { RP-251863}}                                                                                                                     \\ \hline

{\color[HTML]{252525} 20}               & {\color[HTML]{252525} RAN2}                   & {\color[HTML]{252525} Non-Terrestrial Networks (NTN) for Internet of Things (IoT) Phase 4}                                          & {\color[HTML]{252525} { RP-251867}}                                                                                                                     \\ \hline

{\color[HTML]{252525} 20}               & {\color[HTML]{252525} RAN1}                   & {\color[HTML]{252525} Study on \gls{6gr}}                                   & {\color[HTML]{252525} { RP-251881}}                                                                                                                     \\ \hline

\end{tabular}
\end{table*}

\section*{3GPP 5G NR-NTN Overview}
\label{sec:ntn_overview}


In this section we review the \gls{3gpp} 5G \gls{nr}-\gls{ntn} standard with a focus on satellite networks (Sec. General Architecture), and describe the relevant specifications in Rel. 17, 18, 19, and 20 (Secs. 3GPP Release 17, 3GPP Release 18, 3GPP Release 19, 3GPP Release 20 respectively), as summarized in Tab.~\ref{tab:3gpp}.


\subsection*{General Architecture}
\label{sub:architecture}

\paragraph{Elements}
The \gls{ntn} architecture consists of:
\begin{itemize}
    \item A terrestrial \gls{ue}, such as a handheld or \gls{iot} terminal located on Earth.
    \item An aerial/space station. Different types of stations can be considered. Satellites (SAT) provide global connectivity, with \gls{geo} satellites offering continuous wide-area coverage, whereas \gls{leo} and \gls{meo} satellites enable lower-latency services through dynamic constellations. \glspl{hap}, operating in the stratosphere (20 km), enable wide-area, cost-effective coverage, but face challenges related to stabilization and refueling. 
    \glspl{uav}, flying at low altitudes (a few hundred meters), offer flexible, on-demand connectivity for temporary events, emergency response and mobile relaying, but their high propulsion energy consumption imposes power constraints. 
    \item A ground gateway that interconnects the NG-\gls{ran} with the \gls{5gc} via the \gls{ng} interface, and the public Internet via the N6 interface.
    \item A service link from the \gls{ue} to the aerial/space~station.
    \item A feeder link, also referred to as the Satellite Radio Interface (SRI) in satellite networks, i.e., the backhaul connecting the satellite to the ground gateway.
\end{itemize}

\begin{figure*}[t!]
    \centering
    \includegraphics[width=0.8\textwidth]{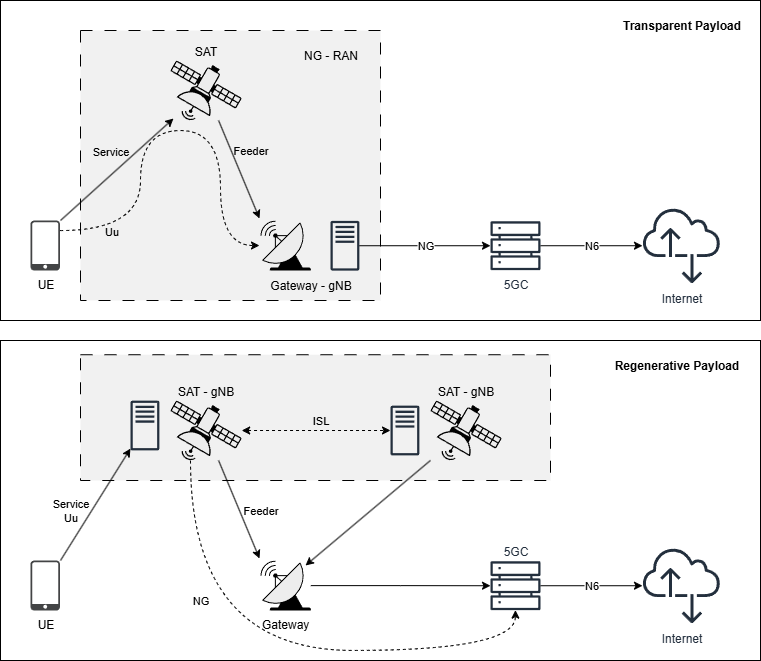}
    \caption{3GPP NTN architectures: transparent (top) and regenerative (bottom). Dashed lines are NR interfaces stretching over multiple~links.}
    \label{fig:fig1}
\end{figure*}

\paragraph{Payload types}
The \gls{3gpp} considers two options:

\begin{itemize}
\item Transparent payload (or ``bentpipe payload'' or ``amplify-and-forward mode''), as illustrated in Fig.~\ref{fig:fig1} (top): the satellite has no onboard processing capabilities. It simply filters, amplifies, and forwards the signal between the service link and the feeder link (and vice versa). In this case, the \gls{gnb} is located on the ground at the gateway, and the satellite functions as an analog RF repeater. In this architecture, the Uu interface terminates at the gateway-\gls{gnb}.

\item Regenerative payload (or ``decode-and-forward mode''), as illustrated in Fig.~\ref{fig:fig1} (bottom): compared to the transparent payload, the satellite is equipped with advanced processing capabilities, including signal regeneration, (de)modulation, decoding, and switching/routing, so it acts as a \gls{gnb} from the sky. 
In this case, the Uu interface is on the service link between the \gls{ue} and the satellite, while the gateway functions as a \gls{tnl} node, providing connectivity between the \gls{ran} (the \gls{gnb}) and the \gls{5gc} components.
This architecture also supports \gls{isl}~connectivity.
\end{itemize}
\paragraph{Channel and antenna}
 The \gls{3gpp} \gls{ntn} channel model is characterized in~\cite{38811}, which is based on the baseline channel model for cellular networks~\cite{38901}. It describes different propagation scenarios for \gls{ntn} (``Dense Urban,'' ``Urban,'' ``Suburban'' and ``Rural''), and provides specific path loss, fast fading and \gls{los} probability models for each of them. Notably, the model accounts for atmospheric absorption based on ITU data, and tropospheric/ionospheric scintillation.
 
Satellite antennas are modeled as circular apertures, while \glspl{ue} employ \gls{upa} antennas, or \gls{vsat} antennas, as specified in \cite{38821}.

\paragraph{Frequency bands}
Satellites have traditionally operated in legacy frequency bands below 6 GHz for wide-area coverage.
Recently, the use of higher frequency bands has been explored, and in some cases adopted, by modern satellite systems, to meet the stringent data rate and low-latency requirements of future wireless services~\cite{giordani2020satellite}.

Reflecting this trend, the \gls{3gpp}, in Rel. 17 and 18~\cite{38101}, has defined several 5G \gls{nr}-\gls{ntn} bands for satellite communications:

\begin{itemize}
    \item L/S (n254: 1610–1626 MHz UL; 2483–2500 MHz~DL);
    \item S (n256: 1980–2010 MHz UL; 2170–2200 MHz DL);
    \item L (n255: 1626.5–1660.5 MHz UL; 1525–1559 MHz DL);
    \item Ka (n510-512: 27.5–30 GHz UL; 17.3–20.2 GHz~DL);
\end{itemize}
Furthermore, the \gls{3gpp} also worked on Ku band support (13.75-14.5 GHz UL; 10.70–12.75 GHz DL) in Rel. 19.
In Sec. End-to-End Evaluation of NTN Scenarios we will analyze the performance of different satellite scenarios as a function of the operating frequency.

\subsection*{3GPP Release 17}
\label{sub:rel17}
\gls{3gpp}'s focus shifted from solely terrestrial networks (pre-Rel. 14, before 2017) to the integration with satellites starting as of Rel. 14 (June 2017). Rel. 15 (June 2019) introduced a flexible architecture to support \glspl{ntn}, while Rel. 16 (July 2020) formally initiated the adaptation of the 5G \gls{nr} physical layer and protocol stack for \gls{ntn} operations.

Rel. 17 (March 2022) marked a major milestone, as it was the first release to standardize direct satellite access communication. The scope included the following assumptions:
\begin{itemize}
    \item Support for {transparent (bentpipe) satellite architectures}, based on both \gls{geo} and \gls{leo} satellites.
    \item Support for use cases like {\gls{embb}} via 5G \gls{nr} and \gls{emtc} for \gls{iot} via  NB-\gls{iot} over satellites.
    \item \glspl{ue} with GNSS capabilities for pre-compensation of timing advance and frequency offset, though this approach is not applicable in scenarios where GNSS reception is poor, e.g., indoors or underground.
    \item \gls{fdd} operation at Frequency Range 1 (FR1), that is at sub-6 GHz, to ensure compatibility with existing terrestrial spectrum allocations and to minimize the impact of Doppler.
    \item Earth-fixed tracking area with Earth-fixed, quasi-Earth-fixed, and Earth-moving cells, to account for \gls{leo} satellites mobility and time-varying coverage~patterns.
\end{itemize}

\subsection*{3GPP Release 18}
\label{sub:rel18}
Rel.~18 (June 2024) built upon Rel.~17 by further enhancing \gls{ntn} access capabilities via the following items:
\begin{itemize}
    \item Support for new frequency bands, including the {n254} band in the L range, and the {n510}, {n511} and {n512} bands in the Ka range at 20-30 GHz, to enable higher data rates.
    \item Enhancements to improve uplink coverage, including \gls{pucch} repetition for more reliable \gls{harq} \gls{ack} signaling, and \gls{pusch} \gls{dmrs} bundling to preserve phase continuity and cope with Doppler shifts caused by satellite mobility.
    \item Enhancements to support network-verified \gls{ue} location based on multiple \gls{rtt} measurements, even with limited or no GNSS availability, assuming a single satellite in view. This approach fulfills the regulatory network requirements for verifying \gls{ue}'s reported location information for emergency and security services.
    \item Enhancements for mobility management, including \gls{ntn}-to-\gls{tn} and \gls{tn}-to-\gls{ntn} mobility, \gls{ntn}-to-\gls{ntn} conditional handover for Earth-moving cells, \mbox{Layer-3} \gls{rach}-free handover, and satellite switch with resynchronization, enabling \glspl{ue} to synchronize with the target satellite without handover to optimize service interruption time, network congestion, and power~efficiency.
\end{itemize}
Beyond focusing on the access, Rel. 18 also addresses satellite backhauling, introducing:
    \begin{itemize}
    \item Mechanisms to handle variable latency and capacity, and packet reordering in the satellite backhaul.
    \item Support for {satellite edge computing}, including onboard {\glspl{upf}}, which enables the satellite to process \gls{ue} traffic locally, rather than routing it through distant terrestrial core networks, to improve latency.
\end{itemize}

\subsection*{3GPP Release 19}
\label{sub:rel19}
Rel. 19, to be closed before the end of 2025 at the time of writing, introduces new Work Items to further improve \gls{ntn} performance with additional services and architectures \cite{rp221169}:
\begin{itemize}
    \item Support for regenerative payloads, including \glspl{gnb} onboard satellites, along with any necessary enhancements to enable seamless intra- and inter-\gls{gnb} mobility.
    \item Downlink coverage enhancements at both FR1 and FR2, including: (i) support for additional reference payload parameters, considering both power-sharing schemes among satellite beams and alternative energy configurations to satisfy satellite power constraints; (ii) optimized deployment strategies that account for \gls{ntn}-specific limitations, such as available power and feeder link bandwidth, ensuring service continuity across the entire satellite footprint while maximizing the overall system throughput; (iii) new evaluation methodologies and relevant Key Performance Indicators (KPIs) for coverage evaluation.
    \item Uplink capacity and cell throughput enhancements at FR1, including the specification of signaling procedures and potential RF requirements to multiplex \glspl{ue} using orthogonal codes when \gls{pusch} repetitions are used. These enhancements also improve signaling efficiency between the CN and NG-RAN.
    \item Signaling of the intended service area for broadcast and multicast services, e.g., Multicast Broadcast Services (MBSs). In particular, this involves specific \gls{sib} signaling to indicate the intended service area in case of a large satellite footprint.
    \item Support for Reduced Capability (RedCap)~\cite{pagin20235g} \glspl{ue} operating at FR1, enabling cost- and energy-efficient \gls{iot} services over satellites. This includes the definition of RF and \gls{rrm} requirements, and specific enhancements to mitigate the issues caused by the \gls{ta} mismatch between the actual \gls{ta} at the \gls{ue} and the estimated \gls{ta} at the \gls{gnb}.
\end{itemize}
\subsection*{3GPP Release 20}
\label{sub:rel20}
Rel. 20 is targeted for March 2027, covering both 5G-Advanced and 6G in the following aspects:

\begin{itemize}
    \item GNSS-resilient 5G-\gls{nr} \gls{ntn} operations. In Rel. 17/18/19, network operations, from uplink time and frequency pre-compensation to location-based conditional handovers, rely on GNSS availability. However, GNSS information at the \gls{ue} may be temporarily unavailable or degraded, e.g., due to GNSS jamming, spoofing, poor satellite geometry, local blockages, solar storms, or sporadic GNSS measurement for power saving~\cite{RP-251863}. Rel. 20 is set to investigate the impact of GNSS interruptions on both initial access and connected mode procedures, and introduce appropriate countermeasures against GNSS spoofing and jamming effects to improve robustness.
    \item Support for \gls{iot} Phase 4. While \gls{ntn}-\gls{iot} was already introduced in \gls{3gpp} Rel. 17 and optimized in Rel. 18/19,  a new Work Item has been launched in Rel. 20 to also support \gls{ims} voice calls over NB-\gls{iot}~\cite{RP-251867}. Specifically, the \gls{3gpp} has the following objectives: (i) to enable semi-persistent scheduling for downlink and uplink voice packet transmissions; (ii)  to update the \gls{rrc} connection setup and emergency call procedures; (iii) to allow \gls{ue} transmit power levels beyond the current PC1 limit, up to 37 dBm, to improve uplink performance; and (iv) to extend \gls{ims} NB-\gls{iot} solutions designed for \gls{geo} satellites to \gls{leo} scenarios with no additional~specifications.
    \item Study new 6G-\gls{ntn} designs. 6G aims at achieving a harmonized radio design that integrates both \glspl{tn} and \glspl{ntn}. Since 6G standardization efforts will influence the whole protocol stack, including waveform, modulation, multiple access, channel coding, antenna, duplex modes, multi-connectivity, mobility, handover, positioning, support for new vertical industries, and deployment scenarios, Rel. 20 has initiated preliminary discussions on a subset of potential directions, starting from 5G gaps and limitations~\cite{rp251881}.
\end{itemize}

\begin{table*}
\caption{Simulation parameters. We consider four representative 3GPP 5G NTN calibration scenarios (SC1, SC4, SC6, SC9) as described in~\cite[Tab. 6.1.1.1-9]{38821} to consider different satellite and frequency configurations. Most of the calibration results in terms of DL \gls{fspl} and \gls{snr} have been already validated with those obtained from the \texttt{ns3-NTN} module in~\cite{sandri23implementation}.}
    \label{tab:parameters}
    \centering
        \begin{tabular}{|l|c|c|c|c|c|c|c|c|}
        \hline
        \multirow{ 2}{*}{\textbf{Parameter}} & \multicolumn{2}{c|}{\begin{tabular}[c]{@{}c@{}}\textbf{LEO-600, S band} \\ \textbf{3GPP SC9} \end{tabular}} & \multicolumn{2}{c|}{\begin{tabular}[c]{@{}c@{}}\textbf{LEO-600, Ka band} \\ \textbf{3GPP SC6} \end{tabular}} & \multicolumn{2}{c}{\begin{tabular}[c]{@{}c@{}}\textbf{GEO, S band} \\ \textbf{3GPP SC4} \end{tabular}} & \multicolumn{2}{|c|}{\begin{tabular}[c]{@{}c@{}}\textbf{GEO, Ka band} \\ \textbf{3GPP SC1} \end{tabular}} \\
        \cline{2-9}
        \multicolumn{1}{|c|}{} & \textbf{Satellite} & \textbf{UE} & \textbf{Satellite} & \textbf{UE} & \textbf{Satellite} & \textbf{UE} & \textbf{Satellite} & \textbf{UE} \\
        \hline
        {Carrier frequency} & \multicolumn{2}{c|}{2 GHz} & \multicolumn{2}{c|}{20 GHz} & \multicolumn{2}{c|}{2 GHz} & \multicolumn{2}{c|}{20 GHz} \\
        \hline
        {Bandwidth} & \multicolumn{2}{c|}{30 MHz} & \multicolumn{2}{c|}{400 MHz} & \multicolumn{2}{c|}{30 MHz} & \multicolumn{2}{c|}{400 MHz} \\
        \hline
        {Elevation angle} & \multicolumn{4}{c|}{30°} & \multicolumn{4}{c|}{12.5°} \\
        \hline
        {DL FSPL} & \multicolumn{2}{c|}{159.1 dB} & \multicolumn{2}{c|}{179.1 dB} & \multicolumn{2}{c|}{190.6 dB} & \multicolumn{2}{c|}{210.6 dB} \\
        \hline
        {DL SNR} & \multicolumn{2}{c|}{6.6 dB} & \multicolumn{2}{c|}{8.5 dB} & \multicolumn{2}{c|}{0 dB} & \multicolumn{2}{c|}{11.6 dB} \\
        \hline
        {Altitude} & 600 km & N/A & 600 km & N/A & 35786 km & N/A & 35786 km & N/A \\
        \hline
        {EIRP} & 34 dBW/MHz & 23 dBm & 4 dBW/MHz & 33 dBm & 59 dBW/MHz & 23 dBm & 40 dBW/MHz & 33 dBm \\
        \hline
        {Antenna diameter} & 2 m & N/A & 0.5 m & 0.6 m & 22 m & N/A & 5 m & 0.6 m \\
        \hline
        {Antenna gain} & 30 dBi & 0 dBi & 38.5 dBi & 39.7 dBi & 51 dBi & 0 dBi & 58.5 dBi & 39.7 dBi \\
        \hline
        {Noise figure} & - & 7 dB & - & 1.2 dB & - & 7 dB & - & 1.2 dB \\
        \hline
    \end{tabular}
\end{table*}

\section*{End-to-End Evaluation of NTN Scenarios}
\label{sec:performance}
In this section we numerically evaluate the performance of different satellite network configurations based on 3GPP 5G NR-NTN specifications, comparing different frequency bands and altitudes. In Sec. Simulation Platform we describe our simulation platform. Then, in Sec. Simulation Results, results are given as a function of different orbit and frequency band parameters.

\subsection*{Simulation Platform}
\label{sub:ns3}
The simulated scenario reflects the \gls{3gpp} calibration specifications from \cite{38821}. It consists of a terrestrial \gls{ue}, wirelessly connected with a satellite at altitude $h$ that provides \gls{gnb} functionalities, and a remote host from which the \gls{ue} is downloading data, all characterized by fixed positions. 
Therefore, we consider a regenerative payload architecture, as described in Sec. General Architecture and considered in \gls{3gpp} Rel. 19.
Specifically, we evaluate the end-to-end throughput, \gls{pdr}, and latency. The latter is defined as the application-to-application delay, encompassing the duration from packet generation at the remote host to  successful packet reception at the \gls{ue}. 
We consider a \gls{leo} satellite at $h=600$ km and a \gls{geo} satellite at $h=35\,786$ km, in both the S and Ka bands, for a total of four representative \gls{3gpp} \gls{ntn} calibration scenarios, as reported in Tab. II.
Specifically, each scenario is characterized by different \gls{fspl} and \gls{snr} regimes.
In particular, the \gls{fspl} is computed using the well-known Friis formula as 
\begin{equation}
    \text{FSPL} = 20\log_{10}(d)+20\log_{10}(f)+92.45,
    \label{eq:fspl1}
\end{equation}
where $d$ is the distance between the satellite and the \gls{ue} in km, and $f$ is the carrier frequency in GHz. The \gls{snr} is computed~as
\begin{equation}
    \text{SNR} = \text{EIRP}+\text{G}/\text{T} - {k} - \text{PL} - {B},
    \label{eq:snr1}
\end{equation} 
where $\text{EIRP}$ is the satellite \acrlong{eirp} expressed in dBW, $\text{G/T}$ is the receiving antenna gain over temperature in dB/K, and $k$ is the Boltzmann constant. The term $\text{PL}$ accounts for several channel attenuation components, including the \gls{fspl} (as per Eq. (1)), atmospheric attenuation, shadowing, scintillation loss, and additional losses, whose values for each calibration scenario can be found in \cite[Tab. 6.1.3.3-1]{38821}. Finally, $B$ is the bandwidth.

The remote host generates packets at a constant source rate $R$, with \gls{udp} as the transport protocol.
We consider a ``Rural'' scenario in \gls{los} condition, and the elevation angles of the \gls{geo} and \gls{leo} satellites are fixed to 12.5\textdegree\, and 30\textdegree, respectively, based on the \gls{3gpp} technical report~\cite{38821}. In contrast to the \gls{3gpp} specification parameters, existing commercial systems consist of dense constellations of satellites so that the elevation angle is generally higher than 70\textdegree\, prior to handover, which ensures more reliable communication performance.
We run simulations in both the S band, that is at \gls{fr1}, and the Ka band, that is at \gls{fr2} in the \gls{mmwave} spectrum. Specifically, we use numerology $\mu=2$ at FR1, corresponding to a subcarrier spacing of 60 kHz, and numerology $\mu=3$ at FR2, corresponding to a subcarrier spacing of 120 kHz, which is consistent with the 5G NR standard specifications.

We conduct simulations in ns-3, one of the most accurate tools for end-to-end network simulations.
Specifically, we use the current version of the \texttt{ns3-NTN} module~\cite{sandri23implementation}, an open-source extension of ns-3, developed to model full-stack satellite communication according to the 3GPP 5G NR-NTN Rel. 17 specifications and beyond.
The module implements several key features: (i) the 3GPP NTN channel model based on~\cite{38811}, thereby including the effects of path loss, atmospheric absorption, scintillation, and fading at different frequencies (in the S, L, and Ka bands); (ii) antenna models based on~\cite{38821}, including circular aperture, VSAT, and UPA antennas; (iii) an NTN-specific Earth-Centered, Earth-Fixed (ECEF) cartesian coordinate system; (iv) accurate modeling of the propagation delay; 
and (v) tailored adjustments to protocol timers (especially for \gls{harq} and \gls{rrc}) to account for the long propagation delay in satellite networks. 
As far as the antenna model is concerned, UPA antennas consist of isotropic elements that radiate uniformly in all directions, while for circular aperture antennas the radiation pattern is expressed as a function of the angle $\theta$ measured from the boresight of the antenna's main beam. Specifically, the antenna gain $G$ is expressed as
\begin{equation}
    \label{eq:circular_aperture}
    G(\theta) = 
    \begin{cases}
        1 & \text{for } \theta = 0; \\
        4\left| \frac{J_1(\kappa a \cdot \sin{\theta})}{ka \cdot \sin{\theta}} \right|^2 & \text{for } 0 < |\theta| \leq 90^{\circ};
    \end{cases}
\end{equation}
where $J_1(x)$ is the Bessel function of the first kind and first order, $a$ is the radius of the antenna's circular aperture, $\kappa = 2f/c$ is the wave number, $f$ is the carrier frequency, and $c$ is the speed of light. Note that $\kappa a$
is equal to the number of wavelengths on the circumference of the aperture, and is independent of the operating frequency.

As specified in \cite{38101}, 3GPP 5G NR-NTN communication at both FR1 and FR2 is primarily defined in \gls{fdd} mode. Since the \texttt{ns3-NTN} simulator is natively based on \gls{tdd}, in this study we focus exclusively on \gls{dl} traffic, so that results do not depend on the underlying duplexing implementation.

\subsection*{Simulation Results}
\label{sub:results}

\begin{figure}[t!]
    \makebox[0pt][l]{\hspace{1.5cm}%
\begin{minipage}[t]{\columnwidth}
    \begin{tikzpicture}[xshift=5cm]

\definecolor{kaBlue}{RGB}{0,114,189} 
\definecolor{kaRed}{RGB}{217,83,25}   
\definecolor{sLightBlue}{RGB}{173,216,230} 
\definecolor{sLightRed}{RGB}{255,182,193} 

\begin{axis}[
    width=0,
height=0,
at={(0,0)},
scale only axis,
xmin=0,
xmax=0,
xtick={},
ymin=0,
ymax=0,
legend style={font=\footnotesize},
ytick={},
    ybar,
    bar width=10pt, 
    legend style={at={(0.8,1.2)}, anchor=south, legend columns=2},
    axis line style={black!50},
    area legend,
    tick style={black},
    tick label style={black},
     column sep=1.5em,
    every node near coord/.style={font=\footnotesize, rotate=90, anchor=west},
]

\addplot+[ybar, bar shift=-5pt, fill=kaBlue, draw=black] coordinates {
(0, 0)
};

\addplot+[ybar, bar shift=5pt, fill=kaRed, draw=black] coordinates {
(0, 0)
};

\addplot+[ybar, bar width=6pt, bar shift=-5pt, draw=black,
    pattern=dots, pattern color=black,
    fill=sLightBlue
] coordinates {
(0, 0)
};

\addplot+[ybar, bar width=6pt, bar shift=5pt, draw=black,
    pattern=dots, pattern color=black,
    fill=sLightRed
] coordinates {
    (0, 0)
};

\legend{Ka band (LEO), Ka band (GEO), S band (LEO),  S band (GEO)}

\end{axis}
\end{tikzpicture}
\end{minipage}}\\
\vspace{0.5cm}
    \begin{minipage}[t]{0.99\columnwidth}
        \centering
       \setlength\fheight{0.7\columnwidth}
    \setlength\fwidth{\columnwidth}
    \begin{tikzpicture}

\definecolor{kaBlue}{RGB}{0,114,189}
\definecolor{kaRed}{RGB}{217,83,25}
\definecolor{sLightBlue}{RGB}{173,216,230}
\definecolor{sLightRed}{RGB}{255,182,193}

\begin{axis}[
    ybar,
    bar width=10pt,
    height=\fheight,
    width=\fwidth,
    ylabel={Throughput [Mbps]},
    xlabel={Source rate $(R)$ [Mbps]},
    symbolic x coords={1, 100, 250, 500, 750, 1000, 1500},
    xticklabels={1, 100, 250, 500, 750, 1000, 1500},
    xtick=data,
    xticklabel style={align=center},
    ytick={0, 100, 200, 300, 400, 500},
    ymin=0, ymax=500,
    ymajorgrids=true,
    grid style=dashed,
    tick label style={font=\footnotesize},
ylabel style={font=\footnotesize},
xlabel style={font=\footnotesize},
legend style={font=\footnotesize},
    legend style={at={(0.5,-0.2)}, anchor=north, legend columns=4},
    nodes near coords={},
    every node near coord/.style={}
]

\addplot+[ybar, bar shift=-5pt, fill=kaBlue, draw=black] coordinates {
    (1, 1)
    (100, 100)
    (250, 250)
    (500, 302.72)
    (750, 302.72)
    (1000, 302.72)
    (1500, 302.72)
};

\addplot+[ybar, bar shift=5pt, fill=kaRed, draw=black] coordinates {
    (1, 1)
    (100, 100)
    (250, 250)
    (500, 434.448)
    (750, 469.936)
    (1000, 469.936)
    (1500, 469.936)
};

\addplot+[ybar, bar width=6pt, bar shift=-5pt, draw=black,
    pattern=dots, pattern color=black,
    fill=sLightBlue
] coordinates {
(1, 1)
(100, 76.432)
(250, 76.432)
(500, 76.432)
(750, 76.432)
(1000, 76.432)
(1500, 76.432)
};

\addplot+[ybar, bar width=6pt, bar shift=5pt, draw=black,
    pattern=dots, pattern color=black,
    fill=sLightRed
] coordinates {
    (1, 1)
    (100, 21.68)
    (250, 21.68)
    (500, 21.68)
    (750, 21.68)
    (1000, 21.68)
    (1500, 21.68)
};


\end{axis}
\end{tikzpicture}
    \end{minipage}\\
    \vspace{0.5cm}
    \begin{minipage}[t]{0.99\columnwidth} 
        \centering
               \setlength\fheight{0.7\columnwidth}
    \setlength\fwidth{\columnwidth}

\begin{tikzpicture}

\definecolor{kaBlue}{RGB}{0,114,189} 
\definecolor{kaRed}{RGB}{217,83,25}   
\definecolor{sLightBlue}{RGB}{173,216,230} 
\definecolor{sLightRed}{RGB}{255,182,193} 

\begin{axis}[
    ybar,
    bar width=10pt, 
    ylabel={PDR},
        height=\fheight,
    width=\fwidth,
    xlabel={Source rate $(R)$ [Mbps]},
    symbolic x coords={1, 100, 250, 500, 750, 1000, 1500},
    xtick={1, 100, 250, 500, 750, 1000, 1500}, 
    xticklabel style={align=center},
    ymin=0, ymax=1,
    ymajorgrids=true,
    grid style=dashed,
        tick label style={font=\footnotesize},
ylabel style={font=\footnotesize},
xlabel style={font=\footnotesize},
legend style={font=\footnotesize},
    legend style={at={(0.5,1.2)}, anchor=north, legend columns=4},
    axis line style={black!50},
    area legend,
    tick style={black},
    tick label style={black},
    every node near coord/.style={font=\footnotesize, rotate=90, anchor=west},
]

\addplot+[ybar, bar shift=-5pt, fill=kaBlue, draw=black] coordinates {
    (1, 1)
    (100, 1)
    (250, 1)
    (500, 0.60)
    (750, 0.40)
    (1000, 0.30)
    (1500, 0.20)
};

\addplot+[ybar, bar shift=5pt, fill=kaRed, draw=black] coordinates {
    (1, 1)
    (100, 1)
    (250, 1)
    (500, 0.87)
    (750, 0.63)
    (1000, 0.47)
    (1500, 0.31)
};

\addplot+[ybar, bar width=6pt, bar shift=-5pt, draw=black,
    pattern=dots, pattern color=black,
    fill=sLightBlue
] coordinates {
    (1, 1)
    (100, 0.76)
    (250, 0.30)
    (500, 0.15)
    (750, 0.10)
    (1000, 0.08)
    (1500, 0.05)
};

\addplot+[ybar, bar width=6pt, bar shift=5pt, draw=black,
    pattern=dots, pattern color=black,
    fill=sLightRed
] coordinates {
    (1, 1)
    (100, 0.22)
    (250, 0.09)
    (500, 0.04)
    (750, 0.03)
    (1000, 0.02)
    (1500, 0.01)
};


\end{axis}
\end{tikzpicture}
    \end{minipage}\\
    \vspace{0.5cm}
    \begin{minipage}[t]{0.99\columnwidth}
        \centering
              \setlength\fheight{0.7\columnwidth}
    \setlength\fwidth{\columnwidth}

\begin{tikzpicture}

\definecolor{kaBlue}{RGB}{0,114,189} 
\definecolor{kaRed}{RGB}{217,83,25}   
\definecolor{sLightBlue}{RGB}{173,216,230} 
\definecolor{sLightRed}{RGB}{255,182,193} 

\begin{axis}[
    ybar,
    bar width=10pt, 
    ylabel={Latency [ms]},
        height=\fheight,
    width=\fwidth,
    xlabel={Source rate  $(R)$ [Mbps]},
    symbolic x coords={1, 100, 250, 500, 750, 1000, 1500},
    xtick=data,
    xticklabel style={align=center},
    ymin=0, ymax=1000,
    ytick={0, 250, ..., 1000},
    ymajorgrids=true,
    grid style=dashed,
        tick label style={font=\footnotesize},
ylabel style={font=\footnotesize},
xlabel style={font=\footnotesize},
legend style={font=\footnotesize},
    legend style={at={(0.5,-0.2)}, anchor=north, legend columns=4},
    axis line style={black!50},
    tick style={black},
    tick label style={black},
    every node near coord/.style={font=\footnotesize, rotate=90, anchor=west},
]

\addplot+[ybar, bar shift=-5pt, fill=kaBlue, draw=black] coordinates {
(1, 4.519639)
(100, 4.484639)
(250, 9.454063)
(500, 133.926063)
(750, 134.023757)
(1000, 134.046063)
(1500, 134.063815)
};

\addplot+[ybar, bar shift=5pt, fill=kaRed, draw=black] coordinates {
(1, 136.501783)
(100, 136.498567)
(250, 153.205063)
(500, 204.874063)
(750, 216.321407)
(1000, 216.307063)
(1500, 216.313919)
};

\addplot+[ybar, bar width=6pt, bar shift=-5pt, draw=black,
    pattern=dots, pattern color=black,
    fill=sLightBlue
] coordinates {
(1, 5.010712)
(100, 352.34214)
(250, 599.38014)
(500, 677.81214)
(750, 703.777555)
(1000, 716.80214)
(1500, 729.867065)
};

\addplot+[ybar, bar width=6pt, bar shift=5pt, draw=black,
    pattern=dots, pattern color=black,
    fill=sLightRed
] coordinates {
(1, 137.278569)
(100, 800.63214)
(250, 872.56014)
(500, 896.25614)
(750, 904.128355)
(1000, 908.23414)
(1500, 912.110732)
};


\end{axis}
\end{tikzpicture}
    \end{minipage}
    \caption{End-to-end throughput, packet delivery ratio, and latency at the application layer vs. the source rate $R$. We focus on a regenerative payload architecture, and consider a LEO satellite at $h=600$ km vs. a GEO satellite at $h=35\,786$ km, in both S and Ka bands.}
    \label{fig:leo_geo_s_Ka}
\end{figure}
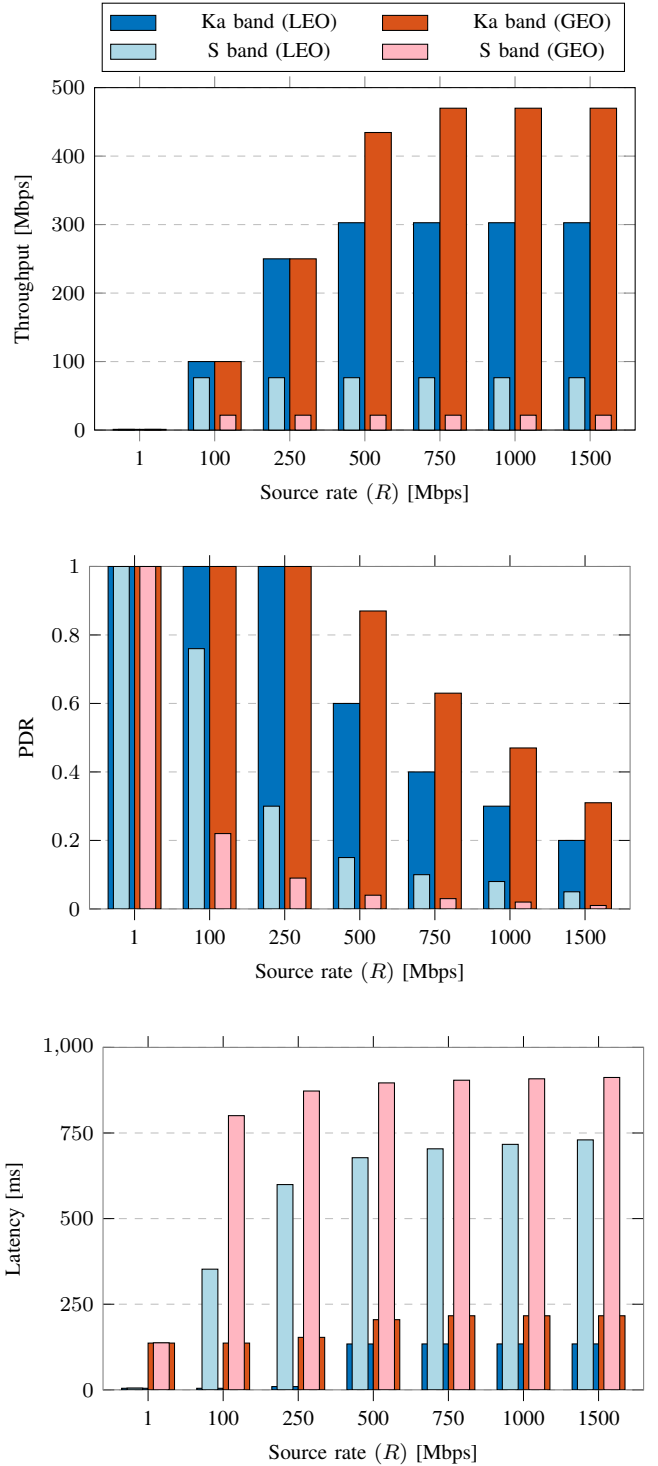

\paragraph*{Satellite orbits}
Focusing on the Ka band, in \cref{fig:leo_geo_s_Ka} (top) we observe that LEO (GEO)  satellites can sustain a source rate at the application up to $R=300$ ($450$) Mbps, which demonstrates the feasibility of NTN communication at different orbits.
After this limit, the network starts to get congested, and the throughput saturates.
Interestingly, despite the longer distance to Earth, the maximum end-to-end throughput for a GEO satellite is up to 50\% higher than for a LEO satellite. 
This difference is primarily due to the lower \gls{snr} for LEO satellites, which is consistent with the 3GPP calibration results in~\cite{38821}. For example, we have that the DL SNR of 3GPP SC1 (GEO) is 11.6 dB, while for SC6 (LEO) it is only 8.5 dB, as reported in Tab. II.
This outcome is in line with realistic satellite design constraints and operational conditions. 
In particular, \gls{geo} satellites can transmit at high power through large circular aperture antennas, providing stable beam pointing and high antenna gain (up to 58.5 dBi, according to the 3GPP parameters in  Tab. II) that can compensate for the severe path loss at large distance. 
In contrast, LEO satellites generally incorporate simpler onboard electronics to reduce hardware and launch costs, and have limited payload capacity, power budget, and antenna gain (only 38.5 dBi, according to the 3GPP parameters in  Tab. II).
Additional offline tests showed that the performance of LEO satellites improves if the \gls{eirp} density is increased beyond the \gls{3gpp} reference value, saturating at 750 Mbps in the Ka band. This confirms that the relative performance of GEO and LEO satellites is primarily determined by the selected simulation parameters, rather than by the orbit itself.

The results are also confirmed by the PDR in \cref{fig:leo_geo_s_Ka} (center).
Notably, we see that GEO satellites experience almost no packet dropping for $R<500$ Mbps, then the PDR decreases to 0.47 when $R=1000$ Mbps, vs. 0.30 in the LEO~scenario.

Conversely, GEO satellites illuminate substantially larger cells on the ground compared to the LEO case, and therefore serve a much higher number of UEs, which may saturate the available channel capacity. 
To better quantify this effect, we evaluate the area capacity density, defined as the network capacity per unit of area. From offline simulations, we observed that the area capacity density is approximately 170 Kbps/km$^2$ for LEO satellites, vs. around 1.4 Kbps/km$^2$ for GEO satellites.

Moreover, \gls{geo} satellites experience significantly higher latency compared to the LEO case due to the longer one-way propagation delay (around 120 vs. 2 ms), as shown in \cref{fig:leo_geo_s_Ka} (bottom).
In general, the latency increases as $R$ increases due to network congestion, causing packet loss.


Notice that, although LEO communication is preferred for latency-sensitive applications like real-time video streaming, as witnessed by several commercial Internet services, the use of GEO satellites remains a viable and often strategic option in certain conditions, e.g., for broadband Internet, given the wide and stable coverage, and the powerful onboard transceiver.


\paragraph*{Frequency bands and bandwidths}
Frequency plays a crucial role in system performance. 
On one side, the maximum bandwidth in the S band is 30~MHz, vs. 400~MHz in the Ka band, which limits the achievable throughput.
On the other side, Ka bands are in general worse from a propagation point of view than the S bands, as the \gls{fspl} is proportional to the carrier frequency. For example, we have that the DL \gls{fspl} of 3GPP SC6 (Ka band) is 179.1 dB, while for SC9 (S band) it is 159.1 dB. As a result, the SNR, and so the overall throughput, deteriorate.
Nevertheless, according to the 3GPP specifications, the UE's and satellite's antennas have around 40 dBi and 10 dBi higher gain, respectively, in the Ka band compared to the S band, as reported in Tab.~\ref{tab:parameters}.
The UE transmission power is also 10 dB higher in the Ka band than in the S band. As a result, the DL SNR of 3GPP SC6 is 8.5 dB, vs. 6.6 dB in SC9. Combined with the larger bandwidth in the Ka band, the channel capacity is around 390~Mbps, vs. 125 Mbps in the S~band.

This performance gap is illustrated in \cref{fig:leo_geo_s_Ka} (top), where the end-to-end throughput in the Ka band  is significantly higher than in the S band. 
Moreover, in the S band the latency increases extremely rapidly with $R$, and it is above 500~ms when $R>250$~Mbps due to network~congestion. 
In any case, satellite communication in the S band may still be desirable in some scenarios, given the more favorable propagation conditions, and better resilience to weather effects and clouds.


\section*{Conclusions}
\label{sec:conclusions}
\glspl{ntn} are emerging as a key enabler to extend 5G connectivity beyond terrestrial infrastructures. Specifically, while satellites offer unique coverage and connectivity advantages, their integration into the 5G ecosystem introduces significant technical challenges due to path loss, latency, and Doppler effects.
Although the 3GPP has introduced formal support for satellite communication in the 5G NR-NTN standard from Rel. 17, existing performance evaluations are often misaligned with standard protocol specifications. To bridge this gap, our study provided an up-to-date overview of recent 3GPP 5G NR-NTN standardization activities through Rel. 20, and evaluated the end-to-end performance of different satellite network architectures through system-level simulations in \mbox{ns-3}.
Simulation results show that LEO satellites outperform GEO satellites in the S band, while it is the opposite in the Ka band. This trend results from the selected simulation parameters, defined according to the 3GPP specifications and calibration scenarios, which model LEO satellites as hardware-constrained platforms, as also discussed in~\cite{wang2021potential}.

Our study focuses on a static, single-UE setup with \gls{udp} traffic, as these assumptions are in line with current 3GPP calibration specifications. As part of our future work, we plan to significantly extend this analysis to more complex and realistic NTN satellite configurations, for example to evaluate the effect of multi-UE scheduling, \gls{tcp} variants, and time-varying LEO geometry with handovers.

\balance
\section*{Code availability}
The underlying code for this study is publicly available in the \texttt{ns3-NTN} repository, and can be accessed via this link: \href{https://gitlab.com/mattiasandri/ns-3-ntn/-/tree/ntn-dev}{https://gitlab.com/mattiasandri/ns-3-ntn}.

\section*{Acknowledgements}
This work was partially supported by the European Union under the Italian National Recovery and Resilience Plan (NRRP) Mission 4, Component 2, Investment 1.3, CUP C93C22005250001, partnership on ``Telecommunications of the Future'' (PE00000001 -- program ``RESTART''). This work was also partially supported by the European Commission through the European Union’s Horizon Europe Research and Innovation Programme under the Marie Skłodowska-Curie-SE, Grant Agreement No. 101129618, UNITE.
The funder played no role in study design, data collection, analysis and interpretation of data, or the writing of this manuscript.

\section*{Author contributions}
MF and FR designed the simulation framework, conducted the simulations, analyzed the data, and wrote Section End-to-End Evaluation of NTN Scenarios. MG conducted a comprehensive literature review for Sections Introduction and 3GPP 5G NR-NTN Overview, wrote the manuscript, and contributed to the results analysis and discussion. AT reviewed the framework, discussed the results, and revised the manuscript. TS, CM, SH, CL, DP, and MZ revised the manuscript and contributed to Secion 3GPP 5G NR-NTN Overview. All authors read and approved the final manuscript.

\section*{Competing interests}
The authors declare no competing interests.

\bibliographystyle{naturemag}
\bibliography{biblio}
\onecolumn
\end{document}